# The solutions of Dirac equation on the hyperboloid under perpendicular magnetic fields

**Duygu Demir Kızılırmak[1] and Şengül Kuru[2]**

[1] Department of Medical Services and Techniques, Ankara Medipol University, Ankara, Turkey
[2] Department of Physics, Faculty of Science, Ankara University, Ankara, Turkey

E-mail: [1] duygudemirkizilirmak@gmail.com, [2] kuru@science.ankara.edu.tr

**Abstract**

In this study, firstly it is reviewed how the solutions of the Dirac-Weyl equation for a massless charge on the hyperboloid under perpendicular magnetic fields are obtained by using supersymmetric (SUSY) quantum mechanics methods. Then, the solutions of the Dirac equation for a massive charge under magnetic fields have been computed in terms of the solutions which were found before for the Dirac-Weyl equation. As an example, the case of a constant magnetic field on the hyperbolic surface for massless and massive charges has been worked out.

Keywords: Dirac equation, Dirac-Weyl equation, supersymmetric methods, hyperbolic surface

## 1. Introduction

While in non-relativistic quantum mechanics the spin-0 particles are described by the Schrödinger equation, in relativistic quantum mechanics the scalar particles are described by Klein-Gordon equation, the spin-½ particles by means of Dirac equation, and massless spin-½ particles by the Dirac-Weyl equation [1, 2]. Here, we will consider both Dirac and Dirac-Weyl equations. But we will see that in this process we must also solve some related effective Schrödinger equations.

Graphene has received much attention in the last years not only in condensed matter but also in theoretical physics [3]. In graphene, charged quasi-particles are described by the Dirac-Weyl equation [4]. Within this context, it is difficult to confine such massless particles in graphene by applying electric fields because of the strong Klein tunnelling. However, this can be achieved by means of magnetic fields. So, it is quite important to know the behaviour of massless charged particles under magnetic fields. Besides graphene, carbon has many allotropes such as graphite, fullerenes, nanotubes, nanoribbons, nano-cones etc. [3]. They have different geometries, so it is quite relevant to adapt the Dirac-Weyl equation to them. Up to now many studies have been done on Dirac-Weyl equation not only in flat space but also in curved spaces [5-17]. Our group has characterized Dirac-Weyl equations on the plane, sphere, cylinder [5-7] and finally on hyperbolic surfaces [8]. We have solved these problems analytically by using simple SUSY methods like factorization or intertwining.

In this work, our aim is to solve analytically the Dirac equation for the massive particles on the hyperbolic surface under perpendicular magnetic fields by means of the solution of massless Dirac-Weyl equations. Hence, we will investigate the relation of the spectrum, ground states, etc. between these two equations and also the importance of mass.

The organization of this paper is as follows. In the second section, the Dirac-Weyl equation on the hyperbolic geometry under magnetic fields is reviewed briefly. In section 3, the bound state solutions of Dirac equation for massive particles under magnetic field on the hyperboloid are found by using the solutions for massless particles. An example is worked out to illustrate these results. Finally, this paper will end with some conclusions.

## 2. Dirac-Weyl equation on the hyperboloid

First, we will review the method to obtain the Dirac-Weyl equation on a surface by restricting that of the flat ambient space as it was done in a preceding paper [8].

## 2.1 Dirac-Weyl equation on the hyperboloid under perpendicular magnetic fields

The Dirac-Weyl equation in (2+1) dimensions in Cartesian coordinates is

$$v_F(\vec{\sigma}\cdot\vec{p})\Phi(x,y,z,t) = i\hbar \frac{\partial \Phi(x,y,z,t)}{\partial t}, \qquad (1)$$

where $\vec{\sigma} = (\sigma_x, \sigma_y, \sigma_z)$ are Pauli matrices, $\vec{p} = -i\hbar(\partial_x, \partial_y, \partial_z) := -i\hbar(\frac{\partial}{\partial x}, \frac{\partial}{\partial y}, \frac{\partial}{\partial z})$ is the three dimensional momentum operator and $v_F$ is an effective Fermi velocity. If there is an external magnetic field, then the interaction between the Dirac electron and this field can be described by the minimal coupling rule where the momentum operator $\vec{p}$ is replaced by $\vec{p} - \frac{q}{c}\vec{A}$. Here, $q = -e$ is the electron charge and $\vec{A}$ is the vector potential. The magnetic field $\vec{B}$ is defined in terms of $\vec{A}$ as usual,

$$\vec{A} = (A_x, A_y, A_z), \qquad \vec{B} = \vec{\nabla} \times \vec{A}. \qquad (2)$$

If the stationary solution $\Phi(x,y,z,t) = \Psi(x,y,z)e^{-\frac{iEt}{\hbar}}$ is replaced in equation (1), we get the time-independent Dirac-Weyl equation

$$v_F\left[\vec{\sigma}\cdot\left(\vec{p} - \frac{q}{c}\vec{A}\right)\right]\Psi(x,y,z) = E\,\Psi(x,y,z). \qquad (3)$$

The two-sheeted hyperboloid directed along the z axis can be expressed by the equation $x^2 + y^2 - z^2 = -r^2$. The relations between hyperbolic $(r, u, \phi)$ and Cartesian $(x, y, z)$ coordinates (in the upper sheet) are given by

$$x = r\sinh u\cos\phi, \quad y = r\sinh u\sin\phi, \quad z = r\cosh u \qquad (4)$$

where $0 < u < \infty$, $0 \leq \phi < 2\pi$ and $0 < r < \infty$ [8].

In order to get the Dirac-Weyl equation on the hyperbolic surface, we take the appropriate Pauli matrices and restrict the equation in the ambient space to that of the surface by using suitable momentum operators [8, 18].

We will assume that the magnetic field is perpendicular to the surface of the hyperboloid, $r = R = constant$, thus, hereafter we use the following notation $\Psi(R, u, \phi) := \Psi(u, \phi)$. The vector potential is chosen in the form $\vec{A} = A(u)\hat{\phi} = A(u)(-\sin\phi, \cos\phi, 0)$. Therefore, this system has rotational symmetry around the z axis and the total angular momentum

$$J_z = -i\hbar\partial_\phi + \frac{\hbar}{2}\sigma_z \qquad (5)$$

commutes with the Hamiltonian: $[H, J_z] = 0$. So, $(H, J_z)$ have common eigenfunctions: $J_z\Psi(u,\phi) = \lambda\hbar\,\Psi(u,\phi)$ where $\lambda$ is half odd number. Then, if we substitute the two-component spinor by

$$\Psi(u,\phi) = N\begin{pmatrix} e^{i(\lambda-\frac{1}{2})\phi}f_1(u) \\ e^{i(\lambda+\frac{1}{2})\phi}f_2(u) \end{pmatrix} \qquad (6)$$

in the eigenvalue equation $H\,\Psi(u,\phi) = \mathcal{E}\,\Psi(u,\phi)$ where $\mathcal{E} = E/\hbar v_F$, we can separate the variable $\phi$ from the Dirac-Weyl equation. Thus, we have only one variable $u$ in the reduced equations. Then, following the steps given in [8]

$$F(u) = \begin{pmatrix} f_1(u) \\ f_2(u) \end{pmatrix} = e^{-\frac{\sigma_y}{2}u}\begin{pmatrix} \psi_1(u) \\ \psi_2(u) \end{pmatrix} \qquad (7)$$

and using the transformation $(\psi_1(u), \psi_2(u))^T = \frac{1}{\sqrt{\sinh u}}(g_1(u), ig_2(u))^T = \frac{1}{\sqrt{\sinh u}}G(u)$, we obtain the eigenvalue matrix differential equation [8]:

$$\begin{pmatrix} 0 & \frac{i}{R}\partial_u + i\left(\frac{\lambda}{R\sinh u} - \frac{q}{c\hbar}A(u)\right) \\ \frac{i}{R}\partial_u - i\left(\frac{\lambda}{R\sinh u} - \frac{q}{c\hbar}A(u)\right) & 0 \end{pmatrix}\begin{pmatrix} g_1(u) \\ ig_2(u) \end{pmatrix} = \mathcal{E}\begin{pmatrix} g_1(u) \\ ig_2(u) \end{pmatrix}. \qquad (8)$$

Remark that the resulting matrix Hamiltonian is Hermitian and $\mathcal{E} = \frac{E}{\hbar v_F}$. It includes the centrifugal term in $\lambda$ and the magnetic potential $A(u)$ through a minimal coupling.

## 2.2 *The solutions of Dirac-Weyl equation by using supersymmetric methods*

Let us define the first order differential operators

$$L^{\pm} = \mp \partial_u + W(u) \tag{9}$$

where $W(u)$ is the superpotential given by

$$W(u) = -\frac{\lambda}{\sinh u} + \frac{qR}{c\hbar}A(u). \tag{10}$$

Notice that $L^{\pm}$ are adjoint operators. Taking into account these definitions, the matrix Hamiltonian can be rewritten in terms of $L^{\pm}$ as

$$\begin{pmatrix} 0 & -iL^+ \\ iL^- & 0 \end{pmatrix} \begin{pmatrix} g_1(u) \\ ig_2(u) \end{pmatrix} = R\mathcal{E} \begin{pmatrix} g_1(u) \\ ig_2(u) \end{pmatrix} \tag{11}$$

thus,

$$L^+ g_2(u) = R\mathcal{E} g_1(u), \qquad L^- g_1(u) = R\mathcal{E} g_2(u). \tag{12}$$

From the above equations, a pair of decoupled second order effective Schrödinger equations is obtained

$$H_1 g_1(u) = L^+ L^- g_1(u) = \varepsilon g_1(u), \tag{13}$$

$$H_2 g_2(u) = L^- L^+ g_2(u) = \varepsilon g_2(u), \tag{14}$$

where $\varepsilon = R^2 \mathcal{E}^2$. These two equations can be expressed together in matrix form:

$$\begin{pmatrix} L^+L^- & 0 \\ 0 & L^-L^+ \end{pmatrix} \begin{pmatrix} g_1(u) \\ ig_2(u) \end{pmatrix} = \varepsilon \begin{pmatrix} g_1(u) \\ ig_2(u) \end{pmatrix}. \tag{15}$$

The effective Hamiltonians $H_1$ and $H_2$ are called partner Hamiltonians

$$H_1 = -\frac{\partial^2}{\partial u^2} + V_1(u), \qquad H_2 = -\frac{\partial^2}{\partial u^2} + V_2(u). \tag{16}$$

The potentials corresponding to these two effective Schrödinger Hamiltonians are defined in terms of $W(u)$,

$$V_1(u) = W(u)^2 - W'(u), \qquad V_2(u) = W(u)^2 + W'(u). \tag{17}$$

These relations show that $H_1$ and $H_2$ are indeed supersymmetric partner Hamiltonians. The following intertwining relations are satisfied between them:

$$H_2 L^- = L^- H_1, \qquad H_1 L^+ = L^+ H_2. \tag{18}$$

Let the discrete spectrum of Hamiltonian $H_1$ be the set of eigenvalues denoted by $\{\varepsilon_{1,n}\}$, $n = 0,1,...$ corresponding to the set of eigenfunctions $\{g_{1,n}\}$. If the ground state of $H_1$ is annihilated by $L^-$

$$L^- g_{1,0} = 0, \tag{19}$$

from (13) the ground state eigenvalue of $H_1$ is $\varepsilon_{1,0} = 0$. We must check that $g_{1,0}$ satisfies suitable boundary conditions and it is square-integrable in $(0, \infty)$. Then, the discrete spectrum of $H_2$ contains the eigenvalues $\{\varepsilon_{2,n-1}\}$ and the normalized eigenfunctions $\{g_{2,n-1}\}$, which can be written as follows:

$$g_{2,n-1}(u) = \frac{1}{\sqrt{\varepsilon_{1,n}}} L^- g_{1,n}(u), \qquad \varepsilon_{1,n} = \varepsilon_{2,n-1}, \quad n = 1,2,.... \tag{20}$$

Finally, the eigenvalues of equation (15) are

$$\varepsilon_0 := \varepsilon_{1,0} = 0, \quad \varepsilon_n := \varepsilon_{1,n} = \varepsilon_{2,n-1}, \quad n = 1,2,.... \tag{21}$$

Taking into account the above results, the eigenfunctions corresponding to the excited energy states of matrix Hamiltonian given by (8) will be

$$G_{\pm,n}(u) = N \begin{pmatrix} \pm g_{1,n}(u) \\ i g_{2,n-1}(u) \end{pmatrix}, \tag{22}$$

Where the signs are for positive or negative energies. Therefore, the eigenvalues of Dirac-Weyl equation ($E = \hbar v_F \mathcal{E}$) are found in terms of the eigenvalues $\varepsilon$ of the partner Hamiltonians given by (16) from the relation $\varepsilon = R^2 \mathcal{E}^2$:

$$\mathcal{E}_{\pm,n} := \frac{E}{\hbar v_F} = \pm \frac{1}{R}\sqrt{\varepsilon_n}, \quad n = 1,2,\dots \ . \tag{23}$$

Remark that the ground state wavefunction and energy can be found from equations (19) and (20).

## 2.3 Solutions of the Dirac-Weyl equation on the hyperboloid for a constant magnetic field

As an example, we can choose the vector potential in the following form

$$A(u) = -\frac{ch}{eR} A_0 \coth u . \tag{24}$$

This vector potential gives rise to a constant magnetic field,

$$B_{u,\phi}(u) = -\frac{B_0}{R^2} \tag{25}$$

where $B_0 = A_0 \left(\frac{ch}{e}\right)$ is a constant. The sign of $B_{u,\phi}(u)$ determines the direction of the magnetic field. This problem can be considered as the Landau system on the hyperboloid for massless charged particles [8].

This vector potential leads to the superpotential

$$W(u) = A_0 \coth u - \lambda \operatorname{cosech} u, \quad A_0 < \lambda . \tag{26}$$

Then, the partner potentials have the form [19]:

$$V_1(u) = A_0^2 + (A_0^2 + \lambda^2 + A_0)\operatorname{cosech}^2 u - \lambda(2A_0 + 1)\coth u \operatorname{cosech} u, \tag{27}$$

$$V_2(u) = A_0^2 + (A_0^2 + \lambda^2 - A_0)\operatorname{cosech}^2 u - \lambda(2A_0 - 1)\coth u \operatorname{cosech} u. \tag{28}$$

These partner potentials are shape invariant under the shift transformation $V_2(u, A_0 + 1) = V_1(u, A_0) + 2A_0 + 1$.

The function $g_{1,0}$ is annihilated by $L^-$: $L^- g_{1,0} = 0$, where $L^-$ has the form

$$L^- = \partial_u - \lambda \operatorname{cosech} u + A_0 \coth u \tag{29}$$

therefore, the zero-energy ground state wavefunction is

$$g_{1,0}(u) = N \left(\tanh \frac{u}{2}\right)^\lambda \frac{1}{(\sinh u)^{A_0}} . \tag{30}$$

This is physically acceptable under the conditions: $A_0 > 0$ and $\lambda - A_0 \geq 0$ [8].

For these partner Hamiltonians ($H_1, H_2$), the eigenvalues have the following from:

$$\varepsilon_0 = \varepsilon_{1,0} = 0, \quad \varepsilon_n = \varepsilon_{1,n} = \varepsilon_{2,n-1} = A_0^2 - (A_0 - n)^2, \quad n = 1,2 \dots [A_0], \tag{31}$$

where if $A_0$ is not integer, $[A_0]$ is the integer part of $A_0$ and in case $A_0$ is integer, $[A_0]$ will be $A_0 - 1$, so that always $[A_0]$ is less than $A_0$. The partner potentials (27)-(28) and their corresponding finite number of energy eigenvalues for $A_0=5$ are shown in Figure 1. The eigenfunctions corresponding to these eigenvalues are given by

$$g_{1,n}(w(u)) = (w-1)^{s^-/2} (w+1)^{-s^+/2} P_n^{(s^- - 1/2, -s^+ - 1/2)}(w(u)), \tag{32}$$

$$g_{2,n}(w(u)) = (w-1)^{(s^- - 1)/2} (w+1)^{-(s^+ - 1)/2} P_n^{(s^- + 1/2, -s^+ + 1/2)}(w(u)). \tag{33}$$

Here $P_n^{(a,b)}(w(u))$ denotes the Jacobi polynomials and $s^- = \lambda - A_0$, $s^+ = \lambda + A_0$, $w(u) = \cosh u$. From the definition (23), the eigenvalues for Dirac-Weyl equation are [8]:

$$\mathcal{E}_{\pm,n} = \pm \frac{1}{R}\sqrt{A_0^2 - (A_0 - n)^2}. \tag{34}$$

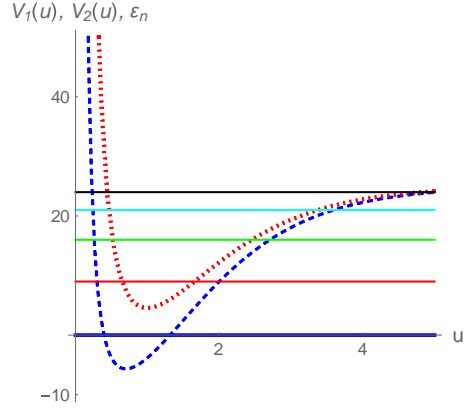

**Fig. 1:** (Colour on-line) Plot of the potentials $V_1$ (dashed line), $V_2$ (dotted line) and the corresponding eigenvalues $\varepsilon_n$ for $n = 0$ (blue, bottom), $n = 1$ (red), $n = 2$ (green), $n = 3$ (cyan), $n = 4$ (black, top). Here, we take $\lambda = 7$ and $A_0 = 5$.

## 3. The Solutions of Dirac Equation on the Hyperboloid

The Dirac equation in (3+1) dimensions for a particle, subject to a magnetic field of vector potential $\vec{A}$, is

$$\left(\beta mc^2 + c\vec{\alpha} \cdot \left(\vec{p} - \frac{e}{c}\vec{A}\right)\right)\Psi(\vec{r},t) = i\hbar \frac{\partial \Psi(\vec{r},t)}{\partial t} \qquad (35)$$

where $m$ denotes its mass, $c$ the speed of light and $e$ its charge. $\beta$ and $\vec{\alpha}$ are matrices defined in terms of Pauli spin matrices $\vec{\sigma}$ as follows [1, 2]

$$\vec{\alpha} = \begin{pmatrix} 0 & \vec{\sigma} \\ \vec{\sigma} & 0 \end{pmatrix}, \qquad \beta = \begin{pmatrix} 1_{2x2} & 0 \\ 0 & -1_{2x2} \end{pmatrix}.$$

We can separate the function $\Psi(\vec{r}, t)$ in time and space coordinates and in terms of two spinors

$$\Psi(\vec{r},t) = e^{-\frac{iEt}{\hbar}} \begin{pmatrix} \varphi_p(\vec{r}) \\ \chi_p(\vec{r}) \end{pmatrix}. \qquad (36)$$

Then, from (35) two coupled $2 \times 2$ matrix equations are obtained

$$c\left(\vec{\sigma} \cdot \left(\vec{p} - \frac{e}{c}\vec{A}\right)\right)\chi_p = (E - mc^2)\varphi_p, \qquad (37)$$

$$c\left(\vec{\sigma} \cdot \left(\vec{p} - \frac{e}{c}\vec{A}\right)\right)\varphi_p = (E + mc^2)\chi_p. \qquad (38)$$

Up to now, the procedure introduced in this section has a general form. Next, we will apply to the Dirac equation its restriction to the hyperboloid by means of hyperbolic coordinates, where we set *r=R=constant* and by making use of momenta tangent to the surface. We also consider magnetic fields with rotational symmetry around the *z*-axis. Hence, we have

$$\varphi_p(u,\phi) = N \frac{e^{-\frac{iu\sigma_y}{2}}}{\sqrt{\sinh u}} \begin{pmatrix} e^{i\left(\lambda - \frac{1}{2}\right)\phi}\varphi_1(u) \\ i\, e^{i\left(\lambda + \frac{1}{2}\right)\phi}\varphi_2(u) \end{pmatrix}, \qquad (39)$$

$$\chi_p(u,\phi) = N \frac{e^{-\frac{iu\sigma_y}{2}}}{\sqrt{\sinh u}} \begin{pmatrix} e^{i\left(\lambda - \frac{1}{2}\right)\phi}\chi_1(u) \\ i\, e^{i\left(\lambda + \frac{1}{2}\right)\phi}\chi_2(u) \end{pmatrix}. \qquad (40)$$

After separation of variables and considering the value $\lambda$ of the total angular momentum $J_z$, Eq. (37) and (38) become:

$$\begin{pmatrix} 0 & -iL^+ \\ iL^- & 0 \end{pmatrix} \begin{pmatrix} \chi_1 \\ i\chi_2 \end{pmatrix} = \frac{R}{c\hbar}(E - mc^2)\begin{pmatrix} \varphi_1 \\ i\varphi_2 \end{pmatrix}, \qquad (41)$$

$$\begin{pmatrix} 0 & -iL^+ \\ iL^- & 0 \end{pmatrix} \begin{pmatrix} \varphi_1 \\ i\varphi_2 \end{pmatrix} = \frac{R}{c\hbar}(E+mc^2)\begin{pmatrix} \chi_1 \\ i\chi_2 \end{pmatrix}. \tag{42}$$

Here, $L^\pm$ are given exactly as in (9) and (10). Then, following the same procedure as in the massless particles, equations (41) and (42) can be decoupled and expressed in a matrix form:

$$\begin{pmatrix} L^+L^- & 0 \\ 0 & L^-L^+ \end{pmatrix}\begin{pmatrix} \varphi_1 \\ i\varphi_2 \end{pmatrix} = \varepsilon \begin{pmatrix} \varphi_1 \\ i\varphi_2 \end{pmatrix}, \qquad \begin{pmatrix} L^+L^- & 0 \\ 0 & L^-L^+ \end{pmatrix}\begin{pmatrix} \chi_1 \\ i\chi_2 \end{pmatrix} = \varepsilon \begin{pmatrix} \chi_1 \\ i\chi_2 \end{pmatrix}, \tag{43}$$

where

$$\varepsilon = \frac{R^2(E^2 - m^2c^4)}{c^2\hbar^2} = R^2\mathcal{E}^2. \tag{44}$$

For each value $\varepsilon$ there are two energy eigenvalues:

$$E_\pm = \pm\sqrt{\frac{c^2\hbar^2}{R^2}\varepsilon + m^2c^4} = \pm\sqrt{c^2\hbar^2\mathcal{E}^2 + m^2c^4}. \tag{45}$$

Thus, in the Dirac eigenvalue problem, the equations (41)-(42) connect the two type of spinors, $\varphi_i$ and $\chi_i$ that compose $\Psi$, while (43) is a consistency equation that must be satisfied by both spinors.

In order to find these Dirac solutions, first we solve equation (43) for the spinor $\chi_i$ (we could as well start with $\varphi_i$). Since this equation coincide with equation (15) of the previous section, we take the same solutions given by the eigenvalues $\varepsilon_n$ of (21) and eigenfunctions $G_{\pm,n}$ of (22). Such a type of solutions $\chi_i$, besides satisfying (43), they also fulfil the Dirac-Weyl equation (11) with eigenvalues

$$R\mathcal{E}_{\pm,n} := \pm\sqrt{\varepsilon_n}, \quad n = 0,1,2,\ldots. \tag{46}$$

Then, from (41) we get:

$$\varphi(u) = \frac{c\hbar\mathcal{E}}{E-mc^2}\chi(u). \tag{47}$$

Next, we replace the solutions $\chi_i$'s into equation (42) to get

$$\chi(u) = \frac{c\hbar\mathcal{E}}{E+mc^2}\varphi(u). \tag{48}$$

Both equations (47) and (48) are consistent.

Then, the spinor solutions $\varphi_i$ and $\chi_i$ of matrix equations (41) and (42) lead to the final solutions of the Dirac equation. The connection of the spinors $\chi_p$ and $\varphi_p$ have the same form:

$$\chi_p(u,\phi) = \frac{c\hbar\mathcal{E}}{E+mc^2}\varphi_p(u,\phi) \tag{49}$$

and

$$\varphi_p(u,\phi) = \frac{c\hbar\mathcal{E}}{E-mc^2}\chi_p(u,\phi). \tag{50}$$

Depending on the sign of the energy eigenvalues, the eigenfunctions will also change according to (11). As we said in the case of Dirac-Weyl equation, there are positive and negative energies $\mathcal{E}_\pm$ and the corresponding Dirac eigenfunctions here are called $\Psi^\pm(u,\phi)$.

The energy eigenvalues of Dirac equation also have two signs and they depend on the square of $\mathcal{E}_\pm$ (so we omit the sub index)

$$E_\pm = \pm\sqrt{c^2\hbar^2\mathcal{E}^2 + m^2c^4}. \tag{51}$$

So, there are two possible values $\mathcal{E}_\pm$ for each $E_\pm$. Thus, we have two independent eigenfunctions corresponding to particles (electrons) for the positive energy $E_+ = +\sqrt{c^2\hbar^2\mathcal{E}^2 + m^2c^4}$:

$$\Psi_{E_+}^+ = \begin{pmatrix} \varphi_p^+ \\ \frac{c\hbar\mathcal{E}_+}{E_++mc^2}\varphi_p^+ \end{pmatrix}, \tag{52}$$

$$\Psi_{E_+}^- = \begin{pmatrix} \varphi_p^- \\ \frac{c\hbar\mathcal{E}_-}{E_++mc^2}\varphi_p^- \end{pmatrix}. \tag{53}$$

A similar situation happens for antiparticles (holes) with negative energy $E_- = -\sqrt{c^2\hbar^2\mathcal{E}^2 + m^2c^4}$:

$$\Psi_{E_-}^+ = \begin{pmatrix} \frac{c\hbar\mathcal{E}_+}{E_--mc^2}\chi_p^+ \\ \chi_p^+ \end{pmatrix}, \tag{54}$$

$$\Psi_{E_-}^- = \begin{pmatrix} \frac{c\hbar\mathcal{E}_-}{E_--mc^2}\chi_p^- \\ \chi_p^- \end{pmatrix}. \tag{55}$$

Here, $\varphi_p^\pm$ and $\chi_p^\pm$ are solutions of the same Dirac-Weyl equation corresponding to $\mathcal{E}_\pm$.

Only one independent function is obtained by taking $E_\pm \to \pm mc^2$ for the ground state energy eigenvalues $E_{0\pm} = \pm mc^2$. We obtain from (52) and (53) the positive energy ground state

$$\Psi_{+mc^2} = \begin{pmatrix} \varphi_0 \\ 0 \end{pmatrix}. \tag{56}$$

From (54) and (55) we have the negative ground state

$$\Psi_{-mc^2} = \begin{pmatrix} 0 \\ \varphi_0 \end{pmatrix} \tag{57}$$

where $\varphi_p^\pm = \chi_p^\pm = \varphi_0$.

### 3.1 Constant magnetic fields

When we choose the vector potential as in (24), the constant magnetic field (25) is obtained. Here, the spinors $\chi_p$ and $\varphi_p$ are the solutions of Dirac-Weyl equation and from (49) and (50) are related by

$$\chi_p(u,\phi) = \pm\sqrt{\frac{E-mc^2}{E+mc^2}}\,\varphi_p(u,\phi), \tag{58}$$

$$\varphi_p(u,\phi) = \pm\sqrt{\frac{E+mc^2}{E-mc^2}}\,\chi_p(u,\phi) \tag{59}$$

and $\varphi_p^\pm(u,\phi)$ have the form

$$\varphi_p^\pm(u,\phi) = N\frac{e^{-\frac{iu\sigma_y}{2}}}{\sqrt{\sinh u}}\begin{pmatrix} \pm e^{i(\lambda-\frac{1}{2})\phi}g_{1,n}(u) \\ i\,e^{i(\lambda+\frac{1}{2})\phi}g_{2,n-1}(u) \end{pmatrix}, \tag{60}$$

where $g_{1,n}(\theta)$ and $g_{2,n-1}(\theta)$ are given by (32) and (33) and the energy is given by (34) and (51). Therefore, the eigenfunctions of Dirac equation also change depending on the sign of the energy eigenvalues: $\Psi^\pm(u,\phi)$. There are two independent eigenfunctions for $E_+ = \sqrt{\frac{c^2\hbar^2}{R^2}(A_0^2 - (A_0-n)^2) + m^2c^4}$:

$$\Psi_{E_+}^+ = \begin{pmatrix} \varphi_p^+ \\ \sqrt{\frac{E_+-mc^2}{E_++mc^2}}\varphi_p^+ \end{pmatrix}, \tag{61}$$

$$\Psi_{E_+}^- = \begin{pmatrix} \varphi_p^- \\ -\sqrt{\frac{E_+-mc^2}{E_++mc^2}}\varphi_p^- \end{pmatrix}, \tag{62}$$

and other two for $E_- = -\sqrt{\frac{c^2\hbar^2}{R^2}(A_0^2 - (A_0-n)^2) + m^2c^4}$ :

$$\Psi_{E_-}^+ = \begin{pmatrix} \sqrt{\frac{E_-+mc^2}{E_--mc^2}}\chi_p^+ \\ \chi_p^+ \end{pmatrix}, \tag{63}$$

$$\Psi_{E_-}^- = \begin{pmatrix} -\sqrt{\frac{E_-+mc^2}{E_--mc^2}}\chi_p^- \\ \chi_p^- \end{pmatrix}. \tag{64}$$

Finally, we have obtained the bound state solutions of Dirac equation for a charge in a constant magnetic field by using the solutions of Dirac-Weyl equation. In figure 2, we plot the energy eigenvalues for Dirac and for Dirac-Weyl Hamiltonians in order to appreciate their differences. It can be seen from the figures, that for the massive case there are negative (corresponding to holes) and positive (corresponding to electrons) eigenvalues starting in the respective ground values $-mc^2$ and $+mc^2$.

There is no bound state with energies in the gap $(-mc^2, mc^2)$. This behavior is opposed to the bound states under an electric field where the energies belong just to that interval. However, for the massless case besides the positive and negative ones there is also the zero-energy state. In the case of electric fields only the zero-energy may have a possible bound state.

The energy levels here obtained may be considered as Landau levels for massless and massive particle cases on the hyperboloid. There are differences with the usual flat Landau levels; for instance, we must remark that in the case of a hyperbolic surface there are only a finite number of eigenvalues (depending on the intensity of the magnetic field) given by formulas (34) and (45). According to formulas (30) and (31) each energy level has infinite degeneracy depending on the angular momentum $\lambda$ that must satisfy an inequality, see (30).

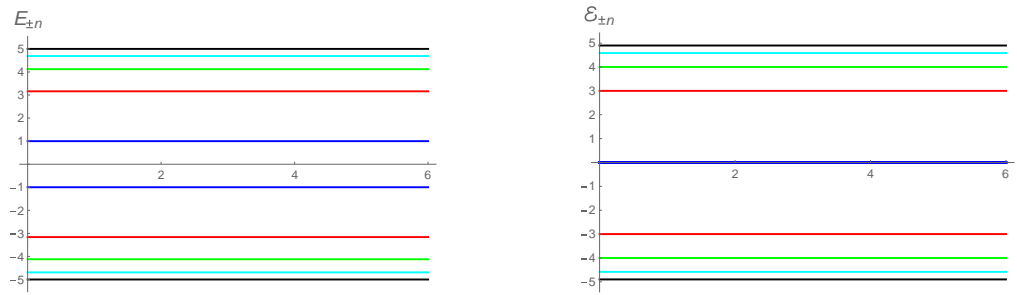

**Fig. 2:** (Colour on-line) Plot of the eigenvalues of Dirac Hamiltonian $E_{\pm,n}$ (left) and Dirac-Weyl Hamiltonian $\mathcal{E}_{\pm,n}$ (right) for $n = 0$ (blue, bottom), $n = 1$ (red), $n = 2$ (green), $n = 3$ (cyan), $n = 4$ (black, top). Here, we choose $A_0 = 5$, $m = 1$ (for massive case), $c = 1$, $\hbar = 1$ (natural units).

## Conclusions

In this work, we revisited the bound state solutions of Dirac-Weyl equation on the hyperbolic surface under perpendicular magnetic fields by using methods of supersymmetry quantum mechanics [8]. This problem could describe the electronic properties of the massless particles on a graphene sheet having the hyperbolic geometry under magnetic fields. When the graphene lattice is subject to strain, its electronic structure will be changed. A type of pseudo magnetic field can be produced by the topology of the strained graphene [21-23]. Then, we obtained the bound state solutions of Dirac equation for massive particles on the hyperboloid under magnetic fields easily from the massless solutions without needing any more computations. By using SUSY methods, the solutions of Dirac equation have been obtained for cylindrical symmetric magnetic fields [20].

In our work, as an example, the case of a constant magnetic field was considered because of its high physical interest. (other magnetic fields can be worked out [5]). This particular problem can be considered as the Landau system on a hyperboloid. As a result, the relation between the solutions and spectrum of the Dirac and the Dirac-Weyl equations have been shown. In figure 2, we plotted these energies in order to compare the eigenvalues of both systems. On the hyperboloid, depending on the intensity of the constant magnetic field, there is a finite number of eigenvalues contrary to the graphene in the flat plane case [5]. From the figures it is seen that for massless particles there is a zero-energy state, but for the massive particles there is not a zero-energy state as it was expected.

Needless to say, this procedure to connect solutions of massless and massive relativistic Dirac equations can be applied to other situations such as in the case of spherical (fullerenes) [6], cylindrical surfaces (nanotubes), or in a flat plane (nano-ribbons).

## Acknowledgements

This work was partially supported by Ankara University BAP No. 20L0430005 and D. Demir Kızılırmak acknowledges Ankara Medipol University. D. Demir Kızılırmak and Ş. Kuru acknowledge J. Negro for his comments and useful discussions.

## References

[1] Schiff I L, *Quantum Mechanics* 1949 (McGraw-Hill Book Company, New York) p. 417
[2] Greiner W, *Relativistic Quantum Mechanics* 1987 (Springer-Verlag, Berlin) p. 447
[3] Castro Neto A H, Guinea F, Peres N M R, Novoselov K S and Geim A K 2009 *Rev. Mod. Phys.* **81** 109
[4] Semenoff G V 1984 *Phys. Rev. Lett.* **53** 2449
[5] Kuru Ş, Negro J and Nieto L M 2009 *J. Phys.: Condens. Matter* **21** 455305
[6] Jakubsky V, Kuru Ş, Negro J and Tristao S 2013 *J. Phys.: Condens. Matter* **25** 165301
[7] Jakubsky V, Kuru Ş, and Negro J 2014 *J. Phys A.: Math. Theor.* **7** 115307
[8] Demir Kızılırmak D, Kuru Ş, and Negro J 2020 *Phys. E Low-Dimens. Syst. Nanostruct.* **118** 113926
[9] Le D-N, Phan A-L, Le V-H and Roy P 2019 *Phys. E Low-dimens. Syst. Nanostruct.* **107** 60
[10] Le D-N, Phan A-L, Le V-H and Roy P 2019 *Phys. E Low-dimens. Syst. Nanostruct.* **114** 113639
[11] Lee H-W, Novikov D S 2003 *Phys. Rev. B,* **68** 155402
[12] Pnueli A 1994 *J. Phys. A Math. Gen.* **27** 1345
[13] Fakhri H 2002 *J. Phys. A Math. Gen.* **35** 6329
[14] Abrikosov A A 2002 *Int. J. Mod. Phys. A* **17** 885
[15] Pudlak M, Pincak R, Osipov V A 2006 *Phys. Rev. B* **74** 235435
[16] Brey L, Fertig H A 2006 *Phys. Rev. B* **73** 235411
[17] Gorbar E V, Gusynin V P 2008 *Ann. Phys.* **323** 2132
[18] Gadella M, Negro J, Nieto L M, Pronko G P and Santander M 2011 *J. Math. Phys.* **52** 063509
[19] Cooper F, Khare A and Sukhatme U 1995 *Phys. Rep.* **251** 267
[20] Contreras-Astorga A, Fernandez C D J and Negro J 2012 *SIGMA* **8** 082
[21] Vozmediano M A H, Katsnelson M I, Guinea F 2010 *Phys. Rep.* **496** 109
[22] Oliva-Leyva M, Naumis G G 2015 *Phys. Lett. A* **379** 2645
[23] Pincak A, Osipov V A 2003 *Phys. Lett. A* **314** 315